\documentclass[10pt,technote]{IEEEtran}
\usepackage{graphicx}
\usepackage{subcaption}
\usepackage{float}
\usepackage{eurosym}

\usepackage[T1]{fontenc}
\begin{document}

\addtolength{\textfloatsep}{-0.3in}
\addtolength{\intextsep}{-0.1in}

\title{
    A Discrete Digital Isolator with Shielded Planar Transformer for Gate Driver Applications
}
\author{
Thomas A. Conway
\thanks{Contact:
Dr. Thomas Conway, ECE Dept, University of Limerick, Ireland.
Email thomas.conway@ul.ie
}
}

\maketitle

\begin{abstract}
  Galvanically isolated gate drivers are required in many power electronics applications.
  This paper describes an implementation of a digital isolator using discrete components with 
  a planar coreless PCB signal transformer providing the isolation. 
  The use of an electrostatic shield between the transformer windings is shown to provide sufficient
  common-mode transient immunity to support IGBT half-bridge drivers, an application subject to a
  severe $dV\over dt$ transient during switching. The design is verified through the experimental
  implementation of a prototype half bridge using two of the proposed discrete digital isolators
  with SiC power transistors and demonstrates common-mode transient immunity greater than 15 kV/$\mu$s
  with very low implementation complexity.
\end{abstract}


\vspace{-0.1in}

\section{Introduction}

  Digital isolators are a common requirement in many power electronics applications to
  galvanically isolate digital control signals from power device gate drivers.
 A wide range of discrete isolated gate driver topologies using planar transformers have been
 reported from applications such as fixed frequency 50\% duty cycle drivers\cite{ref:oldcorless1},
 8\% to 87\% duty cycles\cite{ref:oldcorless2},
 to more recent complex implementations for wider duty cycles such as 15\% to 90\% resonant gate drivers\cite{ref:newGDpowerdrive22}.
 Bidirectional data communications are reported in\cite{ref:recentBidirect26}\cite{ref:recentBidirect26A},
 and\cite{ref:recentBidirectSignPower20}\cite{ref:recentOnedirectSignPower22}\cite{ref:recentBidirectSignPower25} supporting both data communications and gate driver power.

 A challenging  application for these circuits is the high-side driver of a hard switching half-bridge switch topology employing
 IGBTs, MOSFETs both silicon and SiC-bases, or GaN devices.  In these applications, the turn-on and
 turn-off process involves the rapid change in common-mode voltage between the control side
 and the driver side which can result in significant displacement currents through parasitic capacitances
 and result in circuit malfunction.  This effect is commonly parameterized by a common-mode transient immunity
 specification (CMTI) that indicates the maximum common-mode voltage rate of change that can be
 accepted without affecting the data transmission function.

 Earlier circuits employing planar transformers\cite{ref:oldcorless1}\cite{ref:oldcorless2} functioned
 by providing the gate drive power through the planar transformer, and signalled the device turn-on
 by providing power and turn-off by removing power.  This can limit the range of duty cycles
 supported, but reduce the concern of CMTI due to the significant energy required to change the
 switching device's gate capacitance.

 More recent implementations prefer constant driver power and independent data signalling supporting
 full 0 to 100\% duty cycles for modern PWM type applications.   These use some type of ferrite core
 to improve the planar transformer performance\cite{ref:recentBidirect26A}\cite{ref:recentBidirectSignPower25},
 differential signalling\cite{ref:recentBidirect26}\cite{ref:recentOnedirectSignPower22}, and/or
 an electrostatic shield layer\cite{ref:recentBidirect26}\cite{ref:recentBidirect26A}.  Reported
 CMTI specifications are in the range of 15 kV/$\mu$s up to 300 kV/$\mu$s.  However, all result in a moderate 
 to high level of implementation complexity.

 This paper describes a discrete solution employing a coreless planar transformer with
 an electrostatic shield layer implemented using a 4-layer PCB. 
 The design is validated directly in a high-side gate-driver application
 and shown to provide a CMTI specification of 15 kV/$\mu$s.

 The contribution of this work is the demonstration that a practical high-side 
 digital gate drive isolator can be realized using only a coreless PCB transformer,
 single transistor detector and an embedded electrostatic shield. Unlike many reported
 planar transformer isolators that rely on ferrite cores, differential signalling, or complex
 bidirectional communication structures, the proposed approach achieves experimentally 
 verified operation at greater than 15 kV/$\mu$s CMTI while maintaining very low circuit and
 manufacturing complexity.

\begin{figure}[htb]
\begin{center}
    \includegraphics[width=0.48\textwidth]{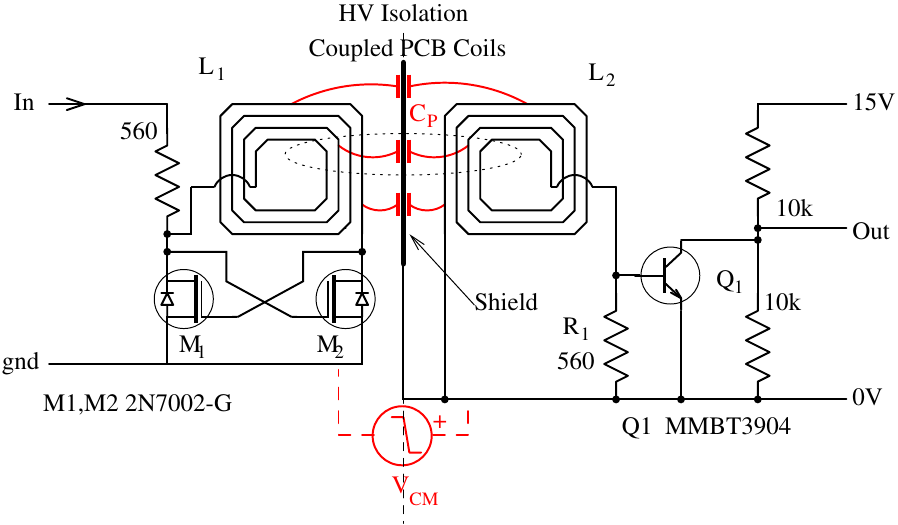}
\caption{Proposed digital isolator with inter-winding shield}
\label{fig:basicschematic}
\end  {center}
\end{figure}

\begin{figure}[htb]
\begin{center}
\begin{subfigure}{0.16\textwidth}
	\includegraphics[width=\columnwidth]{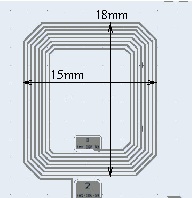}   
	\caption{Layer 4 (bottom)}
    \label{fig:botlayer}
\end{subfigure}
\hfill
\begin{subfigure}{0.16\textwidth}
    \includegraphics[width=\columnwidth]{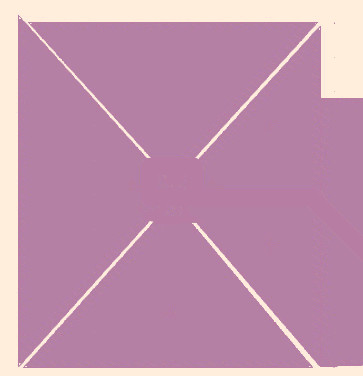}
    \caption{Layer 2}
    \label{fig:secondlayer}
\end{subfigure}\hfill
\begin{subfigure}{0.16\textwidth}
    \includegraphics[width=\columnwidth]{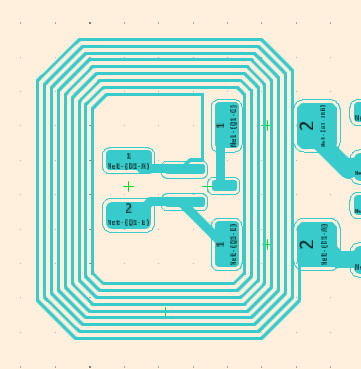}
    \caption{Layer 1 (top)}
    \label{fig:toplayer}
\end{subfigure}
\end  {center}
\caption{PCB layout of coreless transformer and shield}
\label{fig:inductorslayout}
\end{figure}
 
\begin{figure*}[htb]
\begin{center}
\begin{subfigure}{0.32\textwidth}
    \includegraphics[width=\columnwidth]{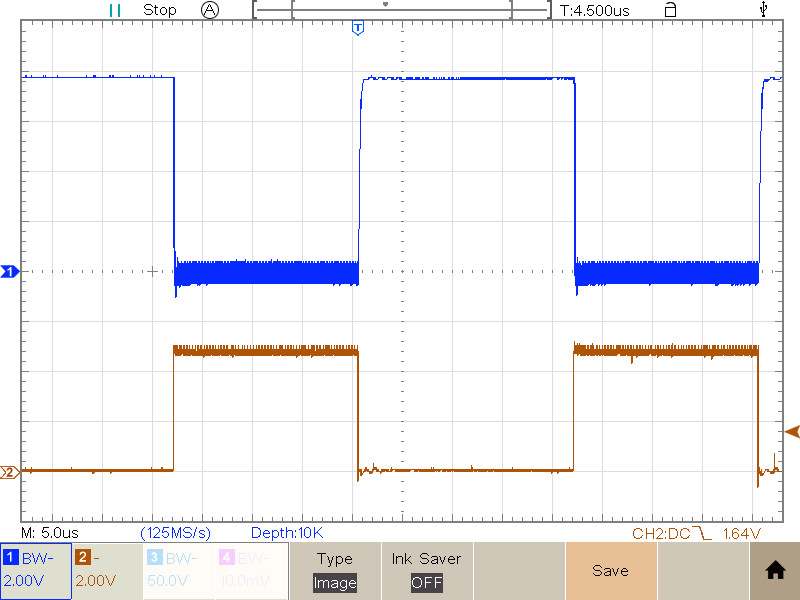}  
    \caption{Functionality}
    \label{fig:isofunc} 
\end{subfigure}
\hfill
\begin{subfigure}{0.32\textwidth}
    \includegraphics[width=\columnwidth]{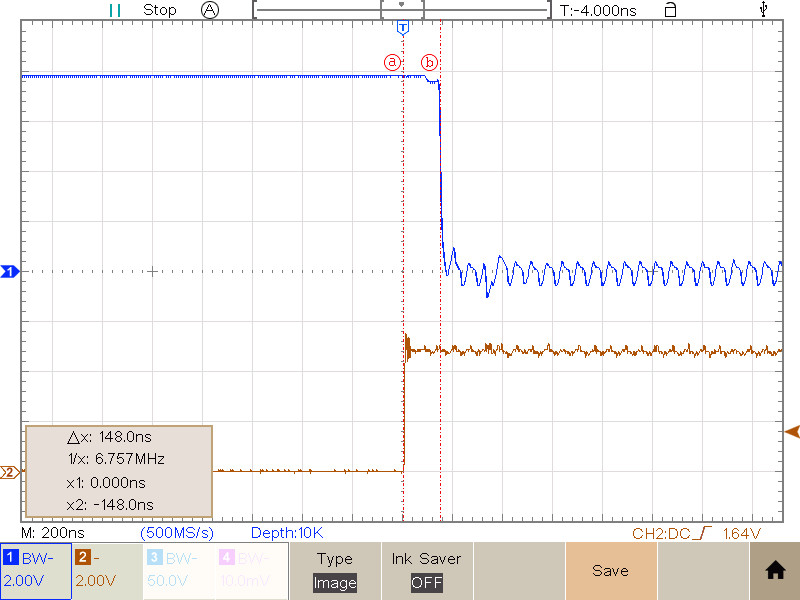}
    \caption{Turn on propagation time}
    \label{fig:isopropl2h}
\end{subfigure}\hfill
\begin{subfigure}{0.32\textwidth}
    \includegraphics[width=\columnwidth]{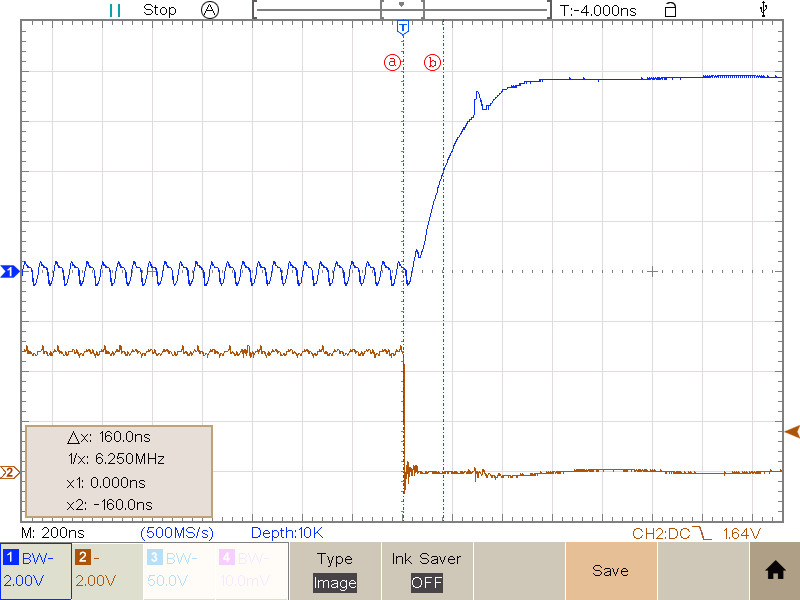}
    \caption{Turn off propagation time}
    \label{fig:isoproph2l}
\end{subfigure}
\end  {center}
    \caption{Functionality of implemented isolator (including shield)}
\label{fig:iosfuncimages}
\end{figure*}

 The paper is structured as follows: section~\ref{sec:propckts} describes the discrete circuit
 implementation of the digital isolator showing both the PCB transformer winding and electrostatic
 shield layout.  The detailed test setup used to evaluate the isolator performance is described
 in section~\ref{sec:testbench} as well as the CMTI results in both the unshielded and shielded
 cases using the high-side driver.  Section~\ref{sec:hbridgetest} shows the operation of the
 complete half-bridge driver with both low and high-side drivers confirming proper operation.
 Section~\ref{sec:rescon} provides a comparison of the proposed implementation with the published literature.

\section{Proposed Isolation Circuits  \label{sec:propckts}}

\begin{figure*}[htb]
\begin{center}
    \includegraphics[width=0.80\textwidth]{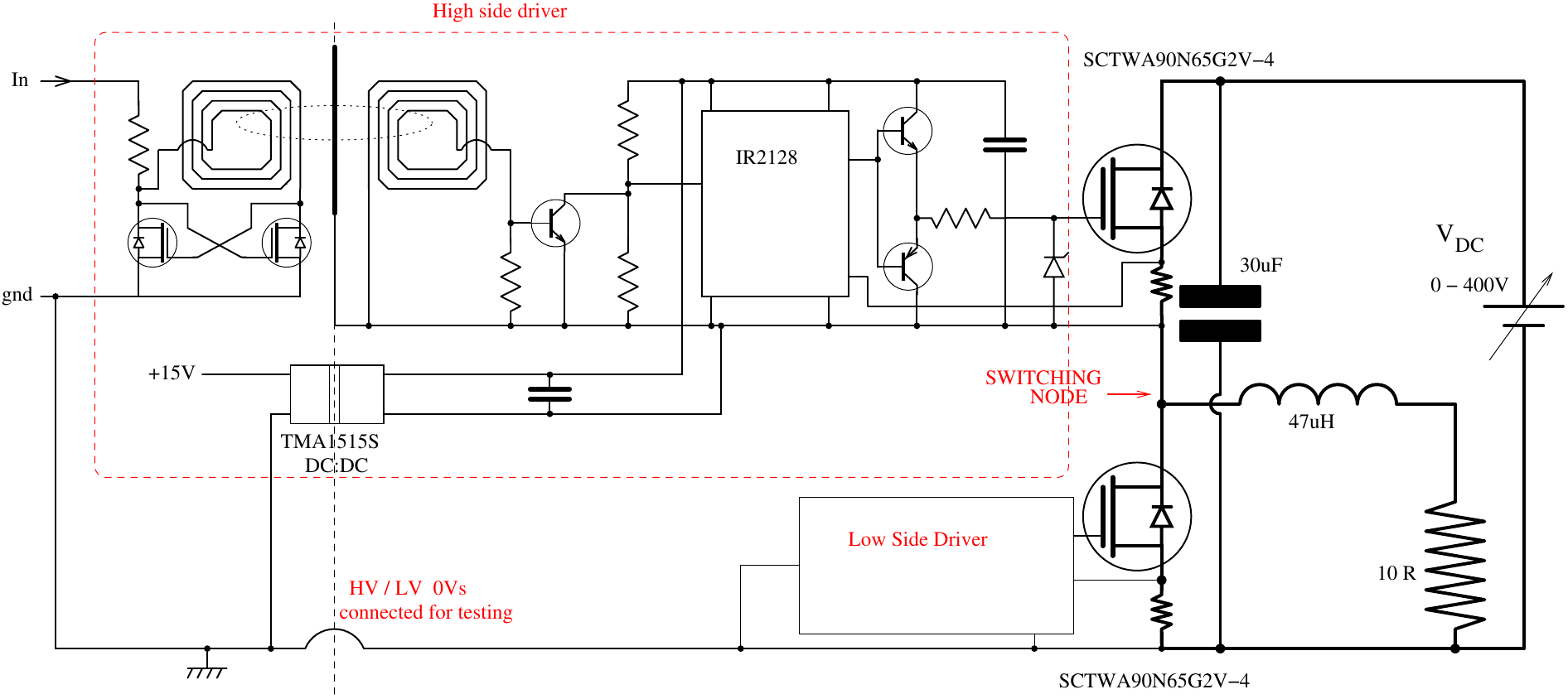}
    \caption{Common mode immunity test circuit using a high-side driver and SiC transistors}
    \label{fig:testbench}
\end  {center}
\end{figure*}

The proposed discrete digital isolator circuit is shown in fig.~\ref{fig:basicschematic}.
The left side is the control side and here is implemented as a classical negative resistance LC oscillator based on a cross coupled
MOSFET pair $M_1$ and $M_2$ with an oscillation frequency around 15 MHz. The internal capacitance
of the transistor pair is sufficient to ensure oscillation in this case.
Alternatively the inductor $L_1$ could be energized with a similar frequency square wave from a modern 
micro-controller port.  

The galvanic isolation functionality is achieved through the use of the coreless PCB transformer,
formed by the bottom side inductor $L_1$ with magnetic coupling to the top side inductor $L_2$. 
Typical PCB materials exhibit dielectric breakdown strengths in 
excess of 20kV/mm\cite{ref:planarxfmer32kV}, 
thus easily providing several kV isolation for the PCB thickness of 1.2 mm used here.
Inductor $L_1$ is implemented on one side of the PCB. 
Fig.~\ref{fig:botlayer} shows the PCB layout of this inductor with a zero ohm surface mount resistor or 
 equivalent used to bridge from the inside to the outside
rather than using vias.  This allows the full insulation properties of the PCB material to
provide the high voltage isolation barrier with sufficient creep distances\cite{ref:tiCreep}.  
The inductor has 8 turns of 200 $\mu$m wide copper trace with 200 um
trace separation and the outside dimensions of 15 mm and 18 mm as shown in the figure.
Using the expressions in\cite{ref:LeeSpiralCalc} yields an estimated inductance of 1.66 $\mu$H, while the measured
inductance on the prototype was 1.7 $\mu$H.

In fig.~\ref{fig:basicschematic}, the right side is the detector structure. 
The inductor $L_2$ is identical to that of $L_1$ but is
placed on the opposite side of the PCB, but directly overlapping it. 
Thus $L_1$ and $L_2$ 
have a significant coupling coefficient and form a 
signal transformer for this application\cite{ref:pcbspiralInd}.

Most magnetically coupled isolators use differential detectors and/or differential transformers 
which result in high complexity circuit implementations. However, the discrete 
detector used here is one of the simplest possible structures consisting of just the bipolar transistor $Q_1$ and the 
resistor $R_1$.
The induced voltage in $L_2$ is effectively half-wave rectified by the bipolar transistor
$Q_1$ with resistor $R_1$ allowing a path for the inductor energy to decay rapidly when
the control signal and corresponding high frequency oscillation stops.  In this circuit
the transistor $Q_1$ is driven into saturation when a positive voltage is induced in $L_2$.
This pulls the output signal active low.  The saturation recovery time of $Q_1$ is sufficient to
keep the output pulled low during the negative cycle of the voltage induced in $L_2$.
This saturation recovery time is typically 100 ns for the bipolar transistor used and effectively
provides glitch filtering functionality when the detector output is active low.

Fig.~\ref{fig:toplayer} shows the PCB layout for $L_2$ which has identical dimensions
to $L_1$.  The resistor $R_1$ and transistor $Q_1$ can be placed inside the spiral inductor
and again zero ohm surface mount resistors or equivalent bridge from the inside to the outside
thus avoiding vias to ensure the full PCB material isolation is available.

The measured waveforms for normal operation of this circuit are shown
in fig.~\ref{fig:iosfuncimages}.  Fig.~\ref{fig:isofunc} shows the proper functionality 
with a 25 kHz input logic signal and the active low output shown. Some small amount of 
the high frequency energizing signal can be seen to feed through due to the half-wave 
rectification, but proper logic levels are readily maintained.  Figs.~\ref{fig:isopropl2h} and
\ref{fig:isoproph2l} show the measured transitions and propagation times demonstrating a 
turn-on delay time of 148 ns and turn-off delay of 160 ns.  
These delays are sufficiently small for typical medium frequency power conversion systems
employing IGBTs or SiC MOSFETs. The similarity of the turn-on and turn-off delays 
make the design suitable for PWM based converters to avoid pulse width distortion.

In these type of  power electronics systems, a key factor is common-mode 
transient immunity (CMTI).  This is particularly true in the case of the high-side 
driver in such an application. The source of the transient is illustrated with the 
fictional voltage source $V_{CM}$ shown in the fig.~\ref{fig:basicschematic} and the
parasitic capacitance between the signal transformer windings represented by $C_P$.
For the prototypes, the measured capacitance was 5 pF.


Consider the controller side (left) being at ground potential and the isolated detector side
being part of a high-side driver that is turned on.  The '0V' of the detector will be
pulled up to the high voltage level of the application.  When the isolator is signalled
to turn-off, the '0V' of the detector will rapidly transition from its high value of typically 
several hundred volts to its low value close to ground potential.  This transition must
happen quickly in hard switched power electronics circuits to minimize switching losses in
the power transistors.  The transition is equivalent to a fast high voltage falling edge 
shown as $V_{CM}$ in the schematic. The net effect can be viewed as a rising edge on the left side
of $C_P$ causing a displacement current flowing into $L_2$, potentially turning
back on transistor $Q_1$.  In section~\ref{sec:unshieldperf}, it is shown that this failure
mechanism occurs and limits the CMTI of the circuit.  The rising edge case is less problematic
due to the inherent holding function provided by the transistor saturation of $Q_1$ when the
isolator output is active low

In order to improve the CMTI of the design, an electrostatic shield is implemented.  
This is achieved by moving to a 4-layer PCB and using one of the internal layers 
as the shield. Fig.~\ref{fig:secondlayer} shows the layout of the electrostatic
shield.  A solid layer would destroy the signal transformer coupling by creating
a shorted turn in the magnetic field path.  Therefore a small gap of 200 $\mu$m is placed
on the four diagonals to reduce magnetically induced eddy currents flowing in the shield 
conductors while minimizing the capacitive coupling between the windings.
The presence of the electrostatic shield reduces the magnetic coupling coefficient
between the windings.  Measurements on prototype planar transformers show a reduction
in the magnetic coupling coefficient from 0.45 with no electrostatic shield down to
0.3 with the electrostatic shield shown.
Although the electrostatic shield reduces the magnetic coupling coefficient, the 
remaining coupling is sufficient to ensure reliable operation of the detector transistor. 
In contrast, the associated reduction in interwinding capacitance significantly 
improves the CMTI performance of the isolator.
The plots in fig.~\ref{fig:iosfuncimages} are for the isolator circuit 
including the electrostatic shield.

   \begin{table*}[htb]
      \begin{center}
      \caption{Comparison of gate drivers utilizing planar transformers}
      \label{tab:comptable}
         \begin{tabular}{|c|c|c|c|c|c|c|}
            \hline
             &  Signal/Power  &  Transformer   & Screen/Shield & Polarity &   CMTI  &  Complexity    \\
            \hline
            \hline
             \cite{ref:oldcorless1} & signal is power 50\% duty cycle & 2-layer PCB coreless & no  & N/A &  N/A&       low  \\
             \cite{ref:oldcorless2} & signal is power 8\% to 87\% duty cycle & 2-layer PCB coreless & no  & N/A &  N/A&       low  \\
             \cite{ref:newGDpowerdrive22} & signal is power 15\%- 90\% duty cycle & six two-layer PCB windings with core &  no & N/A & 35 kV/$\mu$s & moderate \\
             \cite{ref:recentOnedirectSignPower22} & one direction & 3 stacked 2-layer PCBs & no & differential & 100 kV/$\mu$s & moderate \\
             \cite{ref:recentBidirectSignPower25} & bidirectional data and power & 2-layer PCB with ferrite & no & single ended &  100 kV/$\mu$s & moderate \\
             \cite{ref:recentBidirect26A} & bidirectional signal  (CAN) & 4-layer PCB with core & yes & single ended & 125 kV/$\mu$s & high \\
             \cite{ref:recentBidirect26} & bidirectional & 8 layer PCB with core & yes & differential & 300 kV/$\mu$s & high \\
             \cite{ref:recentBidirectSignPower20} &  bidirectional data and power & 6 layer PCB with ferrite core & no & single ended & 15 kV/$\mu$s & moderate \\
             {\it this work } & one direction  &  4-layer PCB coreless  & yes & single ended &  15 kV/$\mu$s     &  very low       \\
           \hline
         \end{tabular}
      \end{center}
   \end{table*}

%
%
\section{Common mode transient immunity evaluation\label{sec:testbench}}

\begin{figure}[htb]
\begin{center}
    \includegraphics[width=0.4\textwidth]{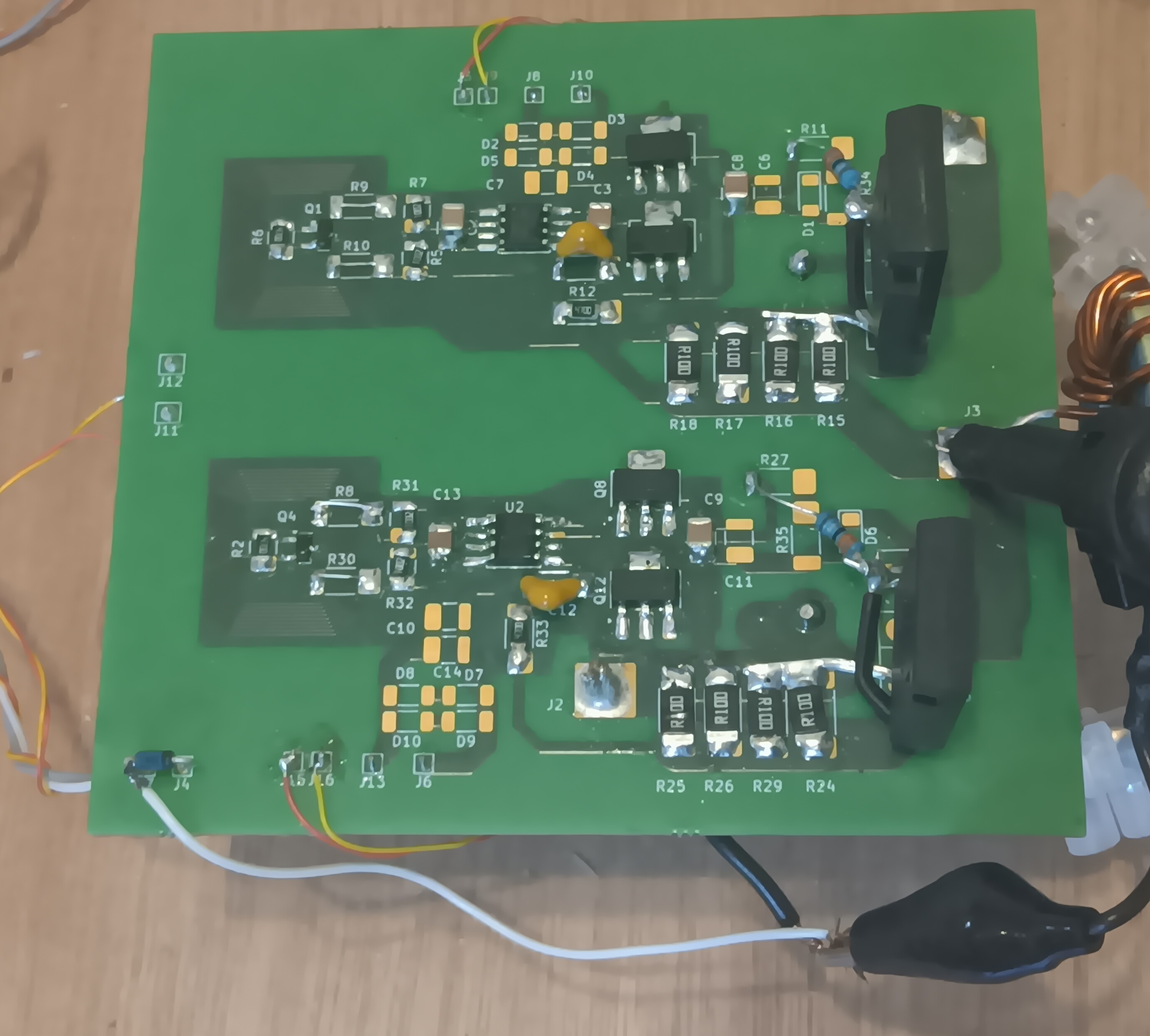}
    \caption{Photo of the half-bridge test PCB}
    \label{fig:boardphoto}   
\end  {center}
\end{figure}

\begin{figure}[htb]
\begin{center}
	\includegraphics[width=0.4\textwidth]{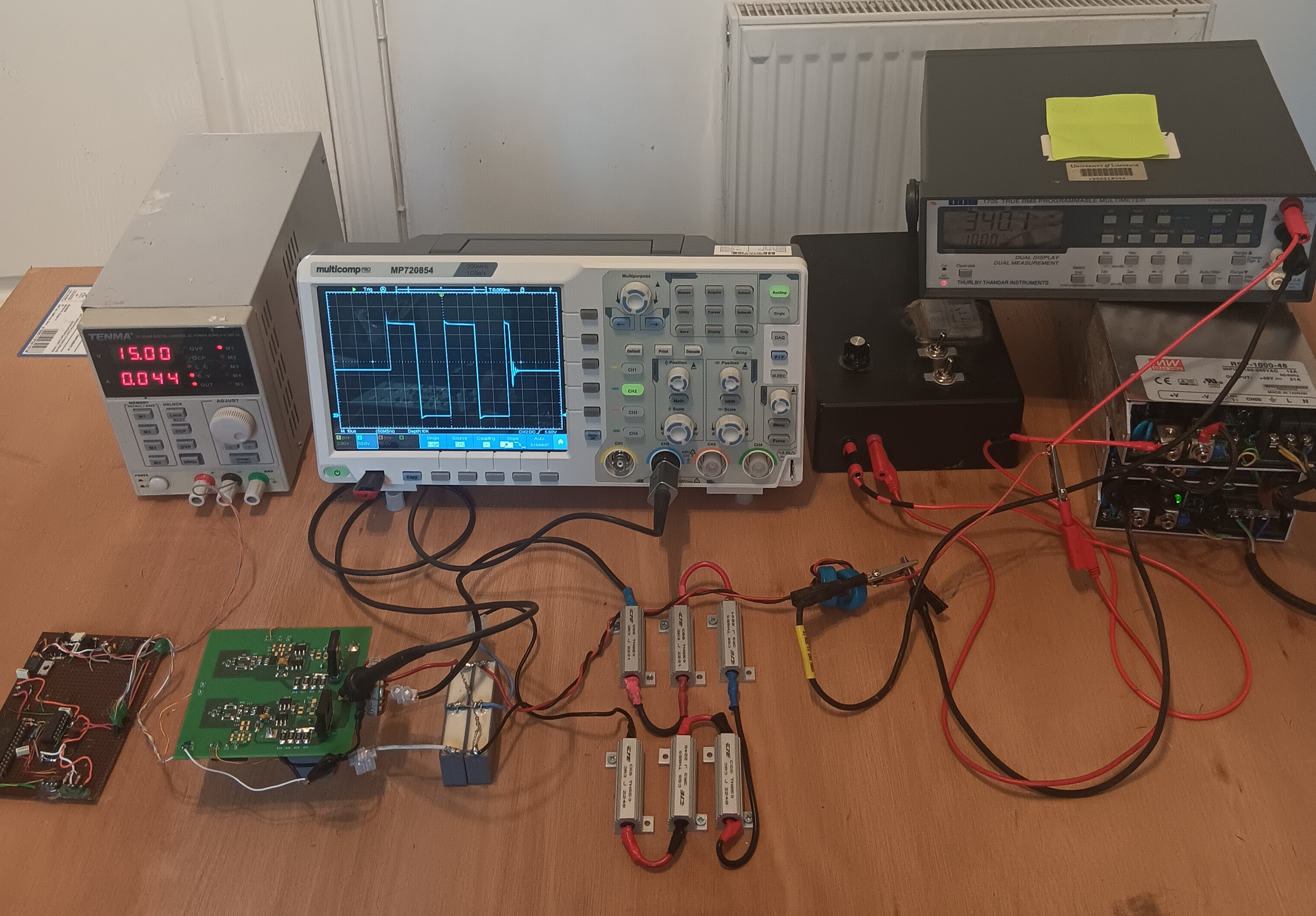}
	\caption{Photo of the complete test system}
	\label{fig:systemphoto}  
\end  {center}
\end{figure}

The most direct method of evaluating the CMTI in the high-side power switching
application is to implement the half-bridge power configuration.  
Fig.~\ref{fig:testbench} shows the test circuit implemented to evaluate the proposed discrete digital
isolator.  A 4-layer PCB implementing two of the isolation circuits from
section~\ref{sec:propckts} together with power drivers, isolated DC:DC converters
and two silicon carbide (SiC) MOSFETs form a half-bridge circuit.  SiC MOSFETs
(SCTWA90N65G2) are selected due to their fast switching times for the purposes of evaluation.
The application of the proposed circuit is intended for the more modest switching times
employed in typical IGBT applications.
Fig.~\ref{fig:boardphoto} shows a photo of the 4-layer test PCB and the spiral
inductors of both the high and low-side isolators can be seen on the left
side of the board.

To evaluate the performance of the isolator, the low-side is held off
and the circuit is used to switch a grounded series inductive and resistive load to the
high voltage DC level $V_{DC}$ variable in this setup from 0 V to 400 V DC.
The test sequence involves a double pulse high-side test, whereby the
high-side is turned on for 19 $\mu$s, providing a positive transition from 0 V to $V_{DC}$
on the switching node and then charging up the load inductor.  
At the end of the pulse period the high-side is turned off causing the switching node
to rapidly transition to below 0 V then turning on the internal diode of the low-side switch.
This sequence is then repeated once more.

Correct functioning of the circuit is verified by varying the DC voltage $V_{DC}$ from 0 V
up to 400 V and observing the switching node for proper operation.
Note that for this test operation, the ground of the controller side is directly connected to the ground
of the high voltage source $V_{DC}$ for testing purposes as this maximizes the common-mode voltage stress
on the high-side driver. 
Fig.~\ref{fig:systemphoto} shows a photo of the complete test bench setup.

\subsection{Measured performance of the unshielded implementation\label{sec:unshieldperf}}

The performance of the discrete digital isolator with no electrostatic shield
was evaluated using the developed test setup. 
As the applied DC voltage $V_{DC}$ exceeded about 200 V, the operation of the 
circuit was observed to fail. 

\begin{figure}[htb]
\begin{center}
    \includegraphics[width=0.8\columnwidth]{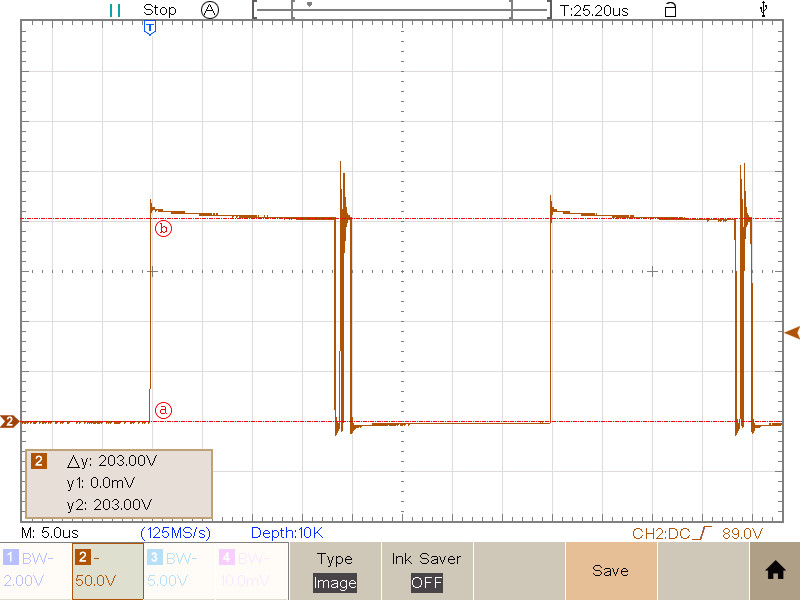}
    \caption{Turn-off failure with no shield}
    \label{fig:det2perffull}
\end  {center}
\end{figure}

Fig.~\ref{fig:det2perffull} shows the observed failure mechanism. At the turn-off
point of the first pulse, the switching node falls as expected, but this transition
which appears as a common-mode transient voltage across the isolator's signal transformer
is now sufficient to cause the isolator's output to turn back on again in error.
The second pulse shows the identical fault occurrence.

\begin{figure}[htb]
\begin{center}
    \includegraphics[width=0.8\columnwidth]{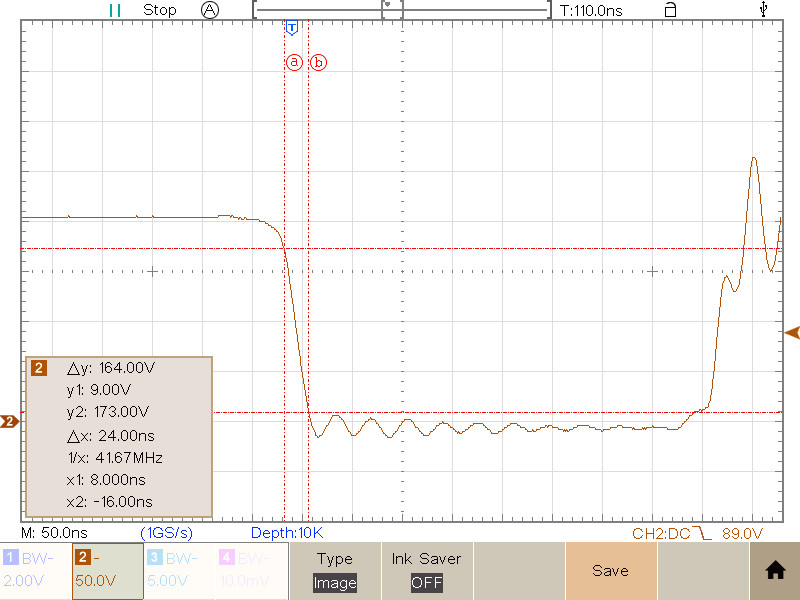}
    \caption{Falling edge time for turn-off failure}
    \label{fig:det2perfzoom}
\end  {center}
\end{figure}

Fig.~\ref{fig:det2perfzoom} shows the high to low transition on the switching node
that triggers this fault.  The measured turn-off transient  shows 
a fall time rate of change of 6.8 kV/$\mu$s which is a modest value of CMTI in the context
of modern digital isolators.

\subsection{Measured performance of the shielded implementation\label{sec:shieldperf}}
\begin{figure}[htb]
\begin{center}
    \includegraphics[width=0.8\columnwidth]{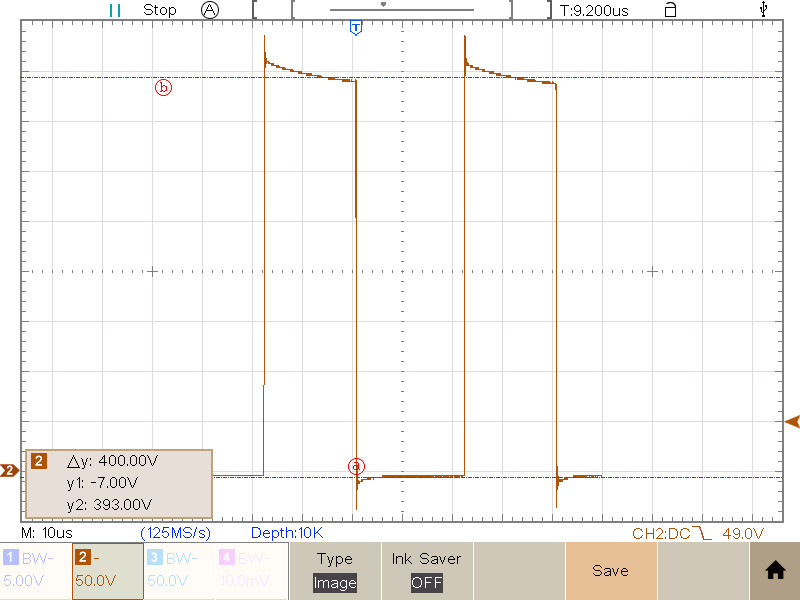}
    \caption{Turn-off correct with shield at 400 V operation}
    \label{fig:det3full}
\end  {center}
\end{figure}

\begin{figure}[htb]
\begin{center}
    \includegraphics[width=0.8\columnwidth]{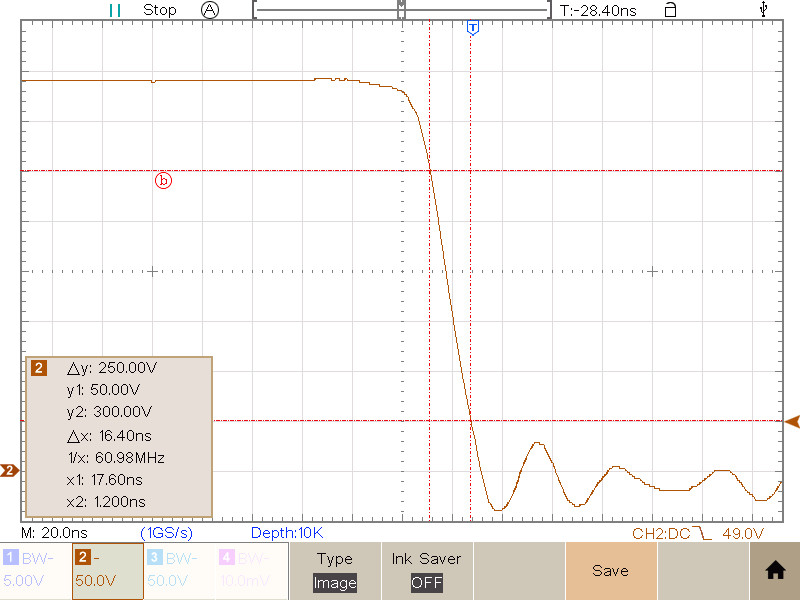}
    \caption{Falling edge time for correct turn-off}
    \label{fig:det3zoom}
\end  {center}
\end{figure}

The performance of the discrete digital isolator with the electrostatic shield
included was then evaluated using the developed test setup.  
In this case, full correct performance was observed over the full applied
voltage range. Fig.~\ref{fig:det3full} shows the double pulse high-side
test with no extraneous switching events present.  Fig.~\ref{fig:det3zoom}
shows the detailed edge transition. The measured turn-off transient  shows
a fall time rate of change of over 15 kV/$\mu$s.

\section{Complete half-bridge operation\label{sec:hbridgetest}}

The  half-bridge configuration with both high-side and low-side drivers
was then tested for normal bridge operation driving the series inductor and
resistor load, but now the other side of the load is connected
to half the $V_{DC}$ supply formed from a capacitor divider. 

\begin{figure}[htb]
\begin{center}
    \includegraphics[width=0.4\textwidth]{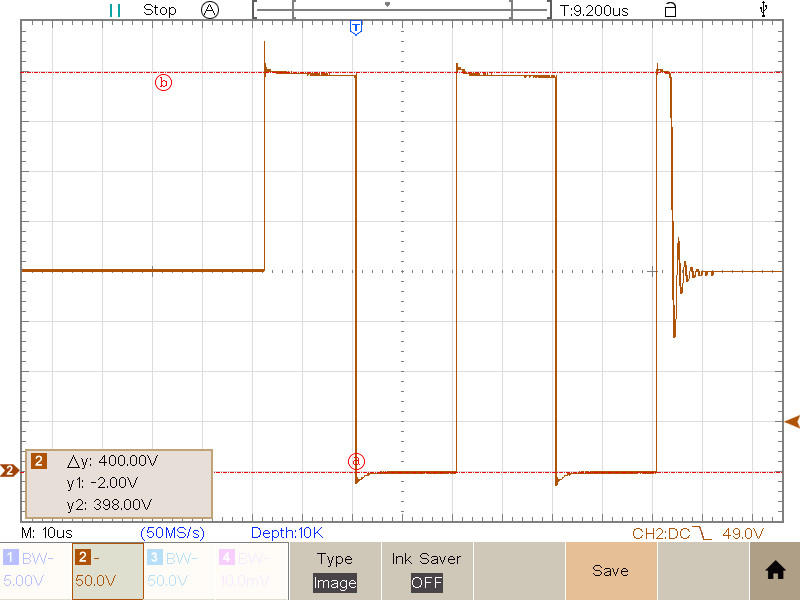}
    \caption{Half bridge with both low and high-side drivers}
    \label{fig:Hbridgescope}
\end  {center}
\end{figure}

Fig.~\ref{fig:Hbridgescope} shows the switching node over a sequence
of both drivers off, one 19 $\mu$s pulse high-side on, 2 $\mu$s dead time, 
one 19 $\mu$s pulse low-side on, 2 $\mu$s dead time, all repeated once
and then both drivers off again.  The figure shows correct operation
of the half-bridge circuit with the 400 V DC supply for this complete sequence.

\section{Conclusions\label{sec:rescon}}
This paper describes a discrete digital isolator suitable for use in 
power electronics applications.  The design uses a coreless planar PCB transformer
structure with an electrostatic shield implemented using a 4-layer PCB and a single
transistor detector structure.  Experimental measurements demonstrate a propagation delay below
200 ns and reliable operation in a SiC MOSFET half-bridge power stage.  
Experimental results demonstrate that the embedded electrostatic shield improves the CMTI
from approximately 6.8 kV/$\mu$s to greater than 15 kV/$\mu$s.

Table~\ref{tab:comptable} shows a comparison of the proposed design with a range
of published gate driver implementations using planar transformers.
The table shows that while the common-mode transient immunity is at the lower 
end of the values listed, the complexity is far less than any of the other designs.

The use of standard multilayer PCB technology, the absence of magnetic cores and the 
minimal component count make the approach attractive for low cost isolated gate
driver systems and other applications requiring robust digital signal isolation.

\end{document}